# Axially Overlapped Multi-Focus Light Sheet with Enlarged Field of View

HONGJIN LI,[1,2] ZIHAN WU,[1] ZHICHAO YANG,[1,2] KARL ZHANGHAO,[1, *] PENG XI[1,3], DAYONG JIN[1,2]

[1] *UTS-SUSTech Joint Research Centre for Biomedical Materials and Devices, Department of Biomedical Engineering, College of Engineering, Southern University of Science and Technology, Shenzhen, Guangdong, China*
[2] *Institute for Biomedical Materials and Devices (IBMD), Faculty of Science, University of Technology Sydney, NSW 2007, Australia*
[3] *Department of Biomedical Engineering, College of Future Technology, Peking University, Beijing, China*

*Corresponding authors: karl.hao.zhang@gmail.com*

**Light sheet fluorescence microscopy provides optical sectioning and is widely used in volumetric imaging of large specimens. However, the axial resolution and the lateral Field of View (FoV) of the system, defined by the light sheet, typically limit each other due to the spatial band product of the excitation objective. Here, we develop a simple multi-focus scheme to extend the FoV, where a Gaussian light sheet can be focused at three or more consecutive positions. Axially overlapped multiple light sheets significantly enlarge the FoV with improved uniformity and negligible loss in axial resolution. By measuring the point spread function of fluorescent beads, we demonstrated that the obtained light sheet has a FoV of 450 μm and a maximum axial FWHM of 7.5 μm. Compared with the conventional single-focus one, the multi-focus Gaussian light sheet displays a significantly improved optical sectioning ability over the full FoV when imaging cells and zebrafish.**

Light sheet fluorescence microscopy (LSFM) has become an indispensable tool in volumetric imaging, with the advances in high spatiotemporal resolution and low photo-toxicity to the fluorescent sample. The open-source design with detailed instructions encourages DIY setups, which has significantly accelerated the wide-range adoption and applications of LSFM. Pioneering works, including OpenSPIM [1], OpenSpin [2], and the recent mesoSPIM [3], provide the detailed protocols for building and using the microscopes. These joint efforts further allow the biology labs to build their LSFM system for a specific application, including more complex schemes, such as multiview excitation or detection configurations [4-6].

The key of LSFM is to generate a thin light sheet with a large field of view (FoV). The use of an illumination objective with a higher N.A. will generate a thinner light sheet, better for improving the axial resolution, as well as protecting the specimen from photobleaching due to the reduced illumination thickness. However, this is at the price of a smaller FoV. To solve this conflict between the axial resolution and FoV, the community has proposed various approaches.

The self-reconstructing beams, such as Bessel beams [7-9], Airy beam [10], attenuation-compensated propagation-invariant beam [11], have been used to maintain the thin sheet over a long propagation distance. Despite their superiority over Gaussian, these beams bring more sidelobes and photobleaching. To minimize the sidelobes, the combinations of light sheet illumination and confocal detection [12, 13], multi-photon excitation [14-17], field modulation [18], phase filters [19], coherent [20-22] and incoherent [23, 24] superpositions have been proposed. Significantly, the dithered lattice light sheet has attracted much attention due to its high resolution and minimal photobleaching [20]. However, there are still different opinions doubting the superiority of self-reconstructing beams over Gaussian beams in light sheet microscopy [25, 26]. Besides, the tradeoff between N.A. and FoV remains to be the bottleneck for large-scale volumetric imaging with high resolution.

The other approach is to generate the light sheet with multiple beams. For example, the mSPIM [27] or SiMView [4] applied the dual-objective excitation configuration to double the FoV with similar axial resolution. Successive focus can be generated by axial scanning with tunable acoustic gradient-index (TAG) [28, 29] lens, electrically tunable lens (ETL) [30, 31], spatial light modulator (SLM) [32, 33], multi-configuration [34] and remote focusing optics [35, 36]. Synchronization with a rolling shutter further increases the axial resolution [35]. However, during the process of creating the virtual thin light sheet, the actual exposure time in the focus region (Rayleigh region) is limited, and specimen beyond the in-focus area suffers from unnecessary photodamage. In a recent effort of the tiled light sheet, thin and small light sheet illumination was spliced to form a large FoV by focus shifting and sequential image acquisition at consecutive positions [32]. However, this approach is at the cost of reduced speed. Besides, all these approaches require precise synchronization, which significantly increases the imaging system's complexity.

Inspired by the remote focusing mechanism [37, 38], here, we design an axially overlapped multi-focus system to break the tradeoff between resolution and FoV, obtaining a high-resolution LSFM with extended FoV. Only one simple component is required: a multi-layer beam splitter (MBS), to generate multiple focal planes at different focal distances. The overlapped light sheets extend the FoV for over an order-of-magnitude, leading to high-resolution LSFM imaging at homogeneous illumination and sharp optical sectioning capability at a large scale. The easy implementation of the MBS makes our LSFM system low-cost, compact, and largely compatible with other LSFM systems. Following the open-source culture, we demonstrate the MF modular design on OpenSPIM, one of the most widely used light sheet system.

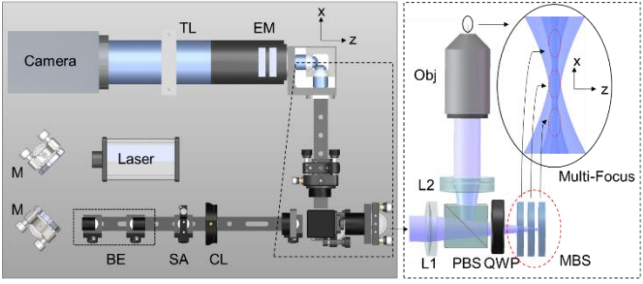

Fig. 1. Schematics of axially overlapped multi-focus light sheet (MF-LS) microscopy. The illumination beam from the laser source (473 nm, bandwidth 0.2 nm, MBL-III-473, CNI) passes through a 3× beam expander, a slit aperture (SA), a cylindrical lens (CL, ACY254-50-A, Thorlabs) and then is guided into the MF component. Finally, three foci are generated after the illumination objective (Obj, Olympus, 10×/N.A. 0.3W). The detailed information of the MF component is illustrated in the right part. It includes one polarization beam splitter (PBS, CCM1-PBS251, Thorlabs), a quarter-wave plate (QWP, WPA4420-450-650, Union Optics), and the multi-layer beam splitter (MBS), laid between a pair of lenses (L1, AC254-50-A, Thorlabs, and L2, AC254-150-A). The detection path, which is composed of one detection objective (Olympus, 40×/N.A. 0.8 W), one emission filter (550LP, Thorlabs), the tube lens (Olympus, 0.5×), and the detector (Hamamatsu ORCA-Flash 4.0), is in the orthogonal direction with a total magnification of 20×.

The axially overlapped multi-focus light sheet (MF-LS) setup integrated into the OpenSPIM system is shown in Fig. 1. The axially overlapped MF-LS can be achieved with one polarization beam splitter (PBS), one quarter-wave-plate (QWP), and MBS. The beam before PBS is horizontally polarized. After passing the PBS and 45° placed QWP, the rays reflected by MBS go through QWP with polarization changed to vertical, and then be reflected by PBS to lens L2 and then the objective. The MBS consists of several beam-splitters and one mirror (Fig. S1), which produces consecutive beams with different divergence, resulting in multiple focusing on the illumination plane. Spacers with variable thickness can adjust the focus drift between neighboring beams, which is proportional to the distance between reflective layers by the ratio of the square of the magnification. Both the reflectivity of each beam-splitter and the distance between neighboring reflectors should be well designed to generate the appropriate light sheet (Supplementary Note 1). This work focuses on the MBS setup with two beam splitters, which produces three primary foci.

We firstly conduct theoretical simulations to study the design and performance of the MBS. Considering Gaussian beam with intensity distribution as follows,

$$I(z,x) = I_0 \left(\frac{w_0}{w(x)}\right)^2 \exp\left(\frac{-2z^2}{w(x)^2}\right) \quad (1)$$

where, $w(x) = w_0 \sqrt{1 + \left(\frac{x}{x_r}\right)^2}$, $x_r = \frac{\pi w_0^2}{\lambda}$, and $w_0 = \frac{n\lambda}{\pi NA}$. $z$ is the radial distance, and $x$ is the axial distance (propagation direction) from the center of the beam, $x_r$ is the Rayleigh length, and $w_0$ is the beam waist. In a first approximation [39], $w_0$ can be calculated as in Eq. (4).

Generally, the in-focus depth of a single Gaussian beam light sheet (FoVs) is considered to be twice the Rayleigh length, with the average axial full-width at half maximum (FWHM) equals to

$$\text{FWHM}_{FOV} = \frac{1}{x_r} \int_0^{x_r} FWHM(x) dx \approx 1.35 w_0 \quad (2)$$

where $FWHM(x) = \sqrt{2\ln 2}\, w(x)$ according to its definition. Thus, the focus drift is also set to be two Rayleigh lengths in the MF model for simplicity (Supplementary Note 2). We use N.A.=0.064 in the simulation, which is consistent with our experimental setting. This N.A. results in a Rayleigh length of ~65 μm, covering FoVs of 130 μm with cellular resolution (FWHM$_{FoVs}$=4.2 μm).

Besides, in our experiments, the axial shift between neighboring beams is much larger than the coherence length of the laser source (Supplementary Note 3). Therefore, the intensity profile of multiple shifted beams would be the sum of each beam under incoherent conditions,

$$I_{AMF}(z,x) = \sum_i I_0 \left(\frac{w_0}{w(x-x_i)}\right)^2 \exp\left(\frac{-2z^2}{w(x-x_i)^2}\right), i=1,2,3 \quad (3)$$

where $x_i$ is the axial shift of each beam.

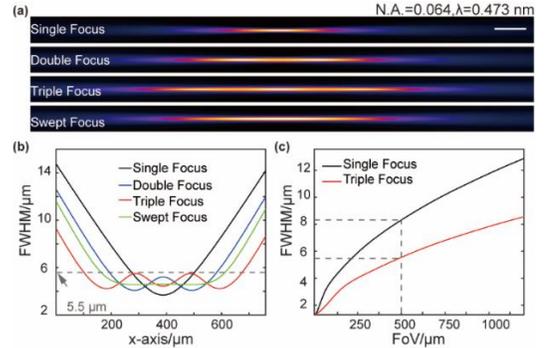

Fig. 2. Principle of axially overlapped MF-LS. (a) simulated image of the excitation beam for single-focus LS, MF-LS, and swept focus LS with the same N.A. (N.A.=0.064). For MF-LS, the same Gaussian beams are focus drifted with doubled Rayleigh lengths (~130 μm). Scale bar: 50 μm. (b) The graph shows the FWHM of the simulated light sheet at different positions along the illumination axis with N.A.=0.064. (c) Both the FoV and FWHM change with the illumination N.A.. The diagram shows the triple-focus LS has a much larger effective FoV than the single-focus LS.

Fig. 2a shows the intensity profiles, and Fig. 2b analyzes the FWHM distribution of the simulated beams under the single-, multi-, and swept focus model with the same N.A.. In the conventional single Gaussian beam, the axial resolution degrades rapidly beyond the Rayleigh region (N.A.$_{ex}$=0.064, $x_r$=65 μm), resulting in a typical FoV of 220 μm with axial resolution better than 5.5 μm, which is the maximum FWHM within the Rayleigh region of triple-focus LS. MF-LS maintains the axial resolution over an enlarged FoV and shows significant improvements in uniformity. Double-focus light sheet can extend the effective FoV to ~380 μm. Triple-focus light sheet further increases the effective FoV to

~570 μm, enlarging the FoV by 2.6 times. Meanwhile, the performance of the MF model is comparable to the swept focus model, showing more advantages at the edge of a larger FoV. Furthermore, we compared the average resolution within the Rayleigh regions between single-focus LS and triple-focus LS in Fig. 2c. When requiring an effective FoV of 500 μm, the average resolution maintains at 5.5 μm in the triple-focus model, while it falls to 8.3 μm in the single-focus model.

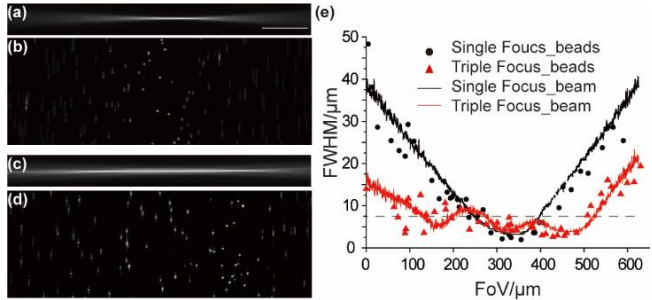

Fig. 3. Experimental measurements of the single-focus LS and triple-focus LS. (a) Profile of line beam imaged in the single-focus model and (c) in the triple-focus model over propagation (from left) length of 665 μm. The focus drift in the triple-focus experiment is around 160 μm. (b) Maximum intensity projection (M. I. P) of microspheres (500 nm) embedded in 2% agarose in the x-z plane under single- and (d) under triple-focus LS illumination. The inserted scale bar is 100 μm. (e) Comparison of the FWHM of the images of the line beam with scattering light (line-plot) and the axial FWHM of fluorescent images of microspheres (dot-plot) over the FoV.

The simulation assumes that multiple Gaussian beams are of the same intensity. To experimentally achieve a uniform light sheet, the reflectivity of every beam splitters should be well designed. With calculation, the optimum reflectivity of two beam splitters should be 23.5% and 39.5% for three foci (Supplementary Note 1). To reduce the cost, we use off-the-shelf beam-splitters with reflectivity of 27% and 40%. Besides the primary focus spots, there also exist some weaker focus spots due to multiple reflected beams, which can also be observed in experimental results. The detection path was set with a magnification of 20× with imaged FoV in terms of the camera sensor area, resulting in a maximum imaging area of 665 × 665 μm². The effective N.A.$_{ex}$ was set to be around 0.064 in the experiments by adjusting the slit aperture (SA) shown in the schematics (Fig. 1). In the MBS, the thickness of spacers is set to be 2 mm, and the thickness of the beam splitters is 3 mm, resulting in the focus shift of ~160 μm, slightly larger than the doubled Rayleigh lengths.

To experimentally implement our MF-LS scheme, firstly, we imaged the line beam with scattered light in the single-focus model (Fig. 3a) and triple-focus model (Fig. 3c), respectively. The cylindrical lens (CL) shown in Fig.1 was temporarily removed, and the SA is replaced with a circular aperture to adjust the effective N.A.$_{ex}$. In the single focus scheme, the laser power was ~1.9 mW in the pupil plane and increased to ~5.4 mW in the triple focus model with an exposure time of 15 ms. From the FWHM distributions of the measured beams (Fig. 3e), it can be seen that, though the single-focus light sheet achieved a higher resolution at the center (~3 μm), the resolution dropped rapidly when the FoV became larger, e.g., ~7.5 μm at the FoV of ~150 μm in diameter. In contrast, in the triple-focus light sheet, the FoV could be extended to approximately ~400 μm in diameter with a consistent resolution better than 7.5 μm.

The calibration experiments with fluorescent microspheres (500 nm) fixed in 2% agarose were carried out under single-focus versus triple-focus illumination, respectively. In the calibration experiment, we changed the detection objective with a low N.A. one (Olympus, 10× /N.A. 0.3 W), in which the axial resolution is determined by the thickness of the light sheet. Fig. 3b shows the Maximum Intensity Projection (M. I. P.) image of the fluorescent beads under single-focus LS illumination, and Fig. 3d shows that under triple-focus LS. The single-focus LS covers a FoV of ~150 μm in diameter with axial resolution beyond 7.5 μm. In contrast, the triple-focus LS covers an FoV of ~ 450 μm, consistent with line beam imaging analysis (Fig. 3e).

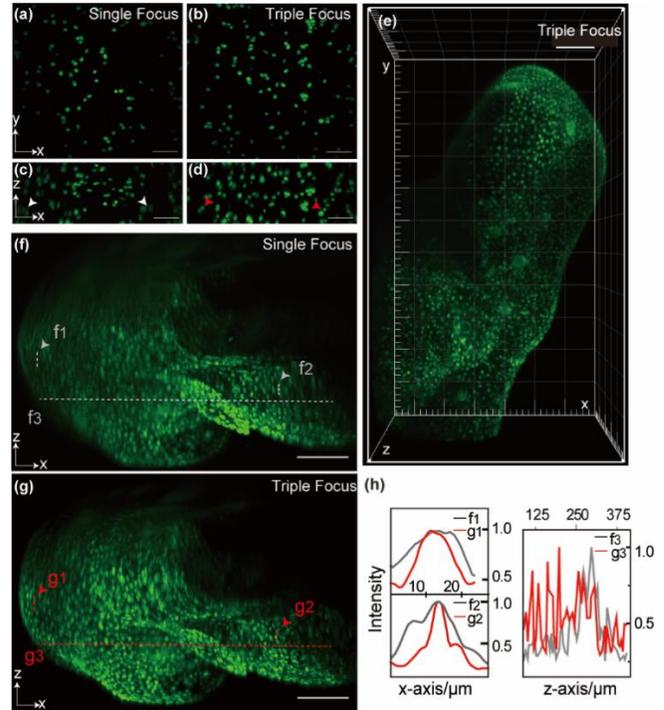

Fig. 4. Experimental data with biological specimens for single-focus (a, c, f) and triple-focus light sheet (b, d, e, g). (a-d) M. I. P of Hela cells labeled by Nile red for single-focus LS (a, c) and triple-focus LS (b, d) in the x-y plane (a, b) and in the x-z plane (c, d). (e) 3D imaging (Δx × Δy × Δz = 665 × 1100 × 460 μm³) of juvenile zebrafish illuminated by MF-LS. (f-g) M. I. P of fixed juvenile Zebrafish labeled SYTOX™ Green Nucleic Acid Stain in the x-z plane (Δx × Δz = 665 × 460 μm²) for triple-focus light sheet (f) and for single-focus light sheet (g). (h) The dot-marked in the M. I. P of biological specimens shows the significant resolution and brightness increment at the edge of the FoV in triple-focus light sheet compared with single-focus light sheet. Scale bar: 100 μm.

Furthermore, we evaluated the performance of MF-LS on different biological specimens. For single focus light sheet illumination, the laser power was set to ~0.06 mW with an exposure time of 25 ms, and then turned up to ~0.18 mW in the triple focus model. Hela cells and juvenile zebrafish were fixed in 2% agarose inside a glass capillary (1 mm diameter) and then pushed out for imaging. Nile-red labeled Hela cells were imaged with single-focus LS and triple-focus LS. The triple-focus illumination showed better illumination uniformity in the x-y perspective (Fig. 4a, b). In the x-z perspective, the cells on the border appear to be elongated and blurred in the single-focus image (Fig. 4c) due to increased FWHM of the light sheet, while those cells are still in round shape in the triple-focus images (Fig. 4d). Afterward, we imaged a volumetric biological specimen. Fixed ~3-day-old Juvenile zebrafish were labeled with SYTOX™ Green Nucleic Acid Stain and imaged. As shown in the x-z M. I. P image in single-focus LS (Fig. 4f), and triple-focus LS (Fig. 4g), the shape of cells was elongated along the z-axis on the border in single-focus LS, showing a larger FWHM than that in triple-focus LS (fig. 4h). Besides, the fluorescent intensity on the border area drops to only 25% of that in the center area in single-focus LS (Fig. 4f, h). In contrast, the fluorescent intensity on the border area remains 50% of that in the center area in triple-focus LS (Fig. 4g, h).

In summary, we propose an axially overlapped multi-focus scheme to break the tradeoff between resolution and FoV of the light sheet. We demonstrate the triple-focus Gaussian light sheet with theoretical simulation and experimental imaging results, exhibiting almost one-order-of-magnitude higher imaging area. The performance of the three-focus light sheet is superior to that of dual-objective excitation with a more simplified system and much lower cost. Moreover, the MF setup is compatible with the dual-objective excitation, which can further extend the effective FoV and make the illumination more uniform when one side of the specimen is heavily scattering. Compared with the tiled light sheet [32], the whole FoV is imaged with only one acquisition so that the imaging speed is not sacrificed.

We anticipate that the axial overlapped MF scheme also applies to four or more foci and multiple excitation objectives. It should be noted that to generate multiple foci, the laser power needs to be increased correspondingly, and it may become harder to keep the parallelism among reflective layers. Thus, novel manufacturing of multiple reflective layers would be necessary. The MF-LS is demonstrated with a medium axial resolution of microns so that the optical aberration is not severe. Obtaining high resolution with large FoV requires not only optical aberration optimization [37, 38], but also novel manufacturing of the MBS. Besides, the coherent interference effect may become a problem with reduced foci shift (Supplementary Note 3). In these applications, the scheme of remote focusing is preferred currently despite the more complicated system and higher cost. Our MF-LS setups are cost-effective, versatile, and can be integrated into many existing light sheet systems with minimum effort. Therefore, our design can upgrade existing light sheet systems that aim at imaging large FoV at cellular resolution.

**Funding.** National Natural Science Foundation of China (62005116, 61729501, 51720105015); Science and Technology Innovation Commission of Shenzhen (KQTD20170810110913065)

**Disclosure.** The authors declare no conflicts of interest.

See Supplement Information for supporting content.

## References

1. P. G. Pitrone, J. Schindelin, L. Stuyvenberg, S. Preibisch, M. Weber, K. W. Eliceiri, J. Huisken, and P. Tomancak, Nat Methods **10**, 598-599 (2013).
2. E. J. Gualda, T. Vale, P. Almada, J. A. Feijo, G. G. Martins, and N. Moreno, Nat Methods **10**, 599-600 (2013).
3. F. F. Voigt, D. Kirschenbaum, E. Platonova, S. Pages, R. A. A. Campbell, R. Kastli, M. Schaettin, L. Egolf, A. van der Bourg, P. Bethge, K. Haenraets, N. Frezel, T. Topilko, P. Perin, D. Hillier, S. Hildebrand, A. Schueth, A. Roebroeck, B. Roska, E. T. Stoeckli, R. Pizzala, N. Renier, H. U. Zeilhofer, T. Karayannis, U. Ziegler, L. Batti, A. Holtmaat, C. Luscher, A. Aguzzi, and F. Helmchen, Nat Methods **16**, 1105-1108 (2019).
4. R. Tomer, K. Khairy, F. Amat, and P. J. Keller, Nat Methods **9**, 755-763 (2012).
5. Y. Wu, P. Wawrzusin, J. Senseney, R. S. Fischer, R. Christensen, A. Santella, A. G. York, P. W. Winter, C. M. Waterman, Z. Bao, D. A. Colon-Ramos, M. McAuliffe, and H. Shroff, Nat Biotechnol **31**, 1032-1038 (2013).
6. R. K. Chhetri, F. Amat, Y. Wan, B. Hockendorf, W. C. Lemon, and P. J. Keller, Nat Methods **12**, 1171-1178 (2015).
7. F. O. Fahrbach, and A. Rohrbach, Opt Express **18**, 24229-24244 (2010).
8. F. O. Fahrbach, P. Simon, and A. Rohrbach, Nature Photonics **4**, 780-785 (2010).
9. J. Yang, L. Gong, Y. Shen, and L. V. Wang, Appl Phys Lett **113**, 181104 (2018).
10. T. Vettenburg, H. I. Dalgarno, J. Nylk, C. Coll-Llado, D. E. Ferrier, T. Cizmar, F. J. Gunn-Moore, and K. Dholakia, Nat Methods **11**, 541-544 (2014).
11. J. Nylk, K. McCluskey, M. A. Preciado, M. Mazilu, Z. Yang, F. J. Gunn-Moore, S. Aggarwal, J. A. Tello, D. E. K. Ferrier, and K. Dholakia, Sci Adv **4**, eaar4817 (2018).
12. F. O. Fahrbach, and A. Rohrbach, Nat Commun **3**, 632 (2012).
13. E. Baumgart, and U. Kubitscheck, Opt Express **20**, 21805-21814 (2012).
14. S. C. Lau, H. C. Chiu, L. Zhao, T. Zhao, M. M. T. Loy, and S. Du, Rev Sci Instrum **89**, 043701 (2018).
15. A. Escobet-Montalban, F. M. Gasparoli, J. Nylk, P. Liu, Z. Yang, and K. Dholakia, Opt Lett **43**, 5484-5487 (2018).
16. N. A. Hosny, J. A. Seyforth, G. Spickermann, T. J. Mitchell, P. Almada, R. Chesters, S. J. Mitchell, G. Chennell, A. C. Vernon, K. Cho, D. P. Srivastava, R. Forster, and T. Vettenburg, Biomed Opt Express **11**, 3927-3935 (2020).
17. F. O. Fahrbach, V. Gurchenkov, K. Alessandri, P. Nassoy, and A. Rohrbach, Opt Express **21**, 13824-13839 (2013).
18. X. Xu, J. Chen, B. Zhang, L. Huang, Y. Zheng, K. Si, S. Duan, and W. Gong, Opt Lett **45**, 4851-4854 (2020).
19. S. Ryu, B. Seong, C. W. Lee, M. Y. Ahn, W. T. Kim, K. M. Choe, and C. Joo, Biomed Opt Express **11**, 3936-3951 (2020).
20. B. C. Chen, W. R. Legant, K. Wang, L. Shao, D. E. Milkie, M. W. Davidson, C. Janetopoulos, X. S. Wu, J. A. Hammer, 3rd, Z. Liu, B. P. English, Y. Mimori-Kiyosue, D. P. Romero, A. T. Ritter, J. Lippincott-Schwartz, L. Fritz-Laylin, R. D. Mullins, D. M. Mitchell, J. N. Bembenek, A. C. Reymann, R. Bohme, S. W. Grill, J. T. Wang, G. Seydoux, U. S. Tulu, D. P. Kiehart, and E. Betzig, Science **346**, 1257998 (2014).
21. T. A. Planchon, L. Gao, D. E. Milkie, M. W. Davidson, J. A. Galbraith, C. G. Galbraith, and E. Betzig, Nat Methods **8**, 417-423 (2011).
22. L. V. Nhu, X. Hoang, M. Pham, and H. Le, Eur Phys J Plus **135** (2020).
23. B. J. Chang, M. Kittisopikul, K. M. Dean, P. Roudot, E. S. Welf, and R. Fiolka, Nat Methods **16**, 235-238 (2019).
24. L. Gao, L. Shao, C. D. Higgins, J. S. Poulton, M. Peifer, M. W. Davidson, X. Wu, B. Goldstein, and E. Betzig, Cell **151**, 1370-1385 (2012).
25. B. J. Chang, K. M. Dean, and R. Fiolka, Opt Express **28**, 27052-27077 (2020).
26. E. Remacha, L. Friedrich, J. Vermot, and F. O. Fahrbach, Biomed Opt Express **11**, 8-26 (2020).
27. J. Huisken, and D. Y. Stainier, Opt Lett **32**, 2608-2610 (2007).
28. W. Zong, J. Zhao, X. Chen, Y. Lin, H. Ren, Y. Zhang, M. Fan, Z. Zhou, H. Cheng, Y. Sun, and L. Chen, Cell Res **25**, 254-257 (2015).
29. K. M. Dean, and R. Fiolka, Opt Express **22**, 26141-26152 (2014).
30. F. O. Fahrbach, F. F. Voigt, B. Schmid, F. Helmchen, and J. Huisken, Opt Express **21**, 21010-21026 (2013).
31. P. N. Hedde, and E. Gratton, Microsc Res Tech **81**, 924-928 (2018).
32. L. Gao, Opt Express **23**, 6102-6111 (2015).
33. L. Gao, W. C. Tang, Y. C. Tsai, and B. C. Chen, Opt Express **27**, 1497-1506 (2019).
34. K. M. Dean, P. Roudot, E. S. Welf, T. Pohlkamp, G. Garrelts, J. Herz, and R. Fiolka, Optica **4**, 263-271 (2017).
35. K. M. Dean, P. Roudot, E. S. Welf, G. Danuser, and R. Fiolka, Biophys J **108**, 2807-2815 (2015).
36. S. Deng, Z. Ding, D. Yuan, M. Liu, and H. Zhou, J Opt Soc Am A Opt Image Sci Vis **38**, 19-24 (2021).
37. E. J. Botcherby, R. Juskaitis, M. J. Booth, and T. Wilson, Opt Lett **32**, 2007-2009 (2007).
38. E. J. Botcherby, R. Juskaitis, M. J. Booth, and T. Wilson, Opt. Commun. **281**, 880-887 (2008).
39. O. E. Olarte, J. Andilla, E. J. Gualda, and P. Loza-Alvarez, Adv Opt Photonics **10**, 111-179 (2018).